\documentclass[10pt,twocolumn,letterpaper]{article}

\usepackage{cvpr}
\usepackage{times}
\usepackage{epsfig}
\usepackage{graphicx}
\usepackage{amsmath}
\usepackage{amssymb}
\usepackage{amsmath,bm}
\usepackage{enumerate}
\usepackage{graphicx}
\usepackage{subfigure}
\usepackage{cite}
\usepackage{caption}
\usepackage{booktabs}
\usepackage{array}
\usepackage{bigstrut,bigdelim,multirow}
\usepackage{mathrsfs}
\usepackage{txfonts}
\usepackage{amsfonts,amssymb}
\usepackage{makecell}
\usepackage{epstopdf}
\usepackage{epsfig}
\usepackage{algorithm}
\usepackage{algorithmic}
\usepackage{hhline}
\usepackage[breaklinks=true,bookmarks=false]{hyperref}

\cvprfinalcopy % *** Uncomment this line for the final submission

\def\cvprPaperID{****} % *** Enter the CVPR Paper ID here

\begin{document}

%%%%%%%%% TITLE
\title{Deep Cross-media Knowledge Transfer}

\author{Xin Huang and Yuxin Peng\thanks{Corresponding author.}\\
Institute of Computer Science and Technology, \\
Peking University, Beijing 100871, China\\
{\tt\small huangxin\_14@pku.edu.cn, pengyuxin@pku.edu.cn}
%\and
}

\maketitle
\pagestyle{empty}  % no page number for the second and the later pages  
\thispagestyle{empty} % no page number for the first page  
%\thispagestyle{empty}

%%%%%%%%% ABSTRACT
\begin{abstract}
   Cross-media retrieval is a research hotspot in multimedia area, which aims to perform retrieval across different media types such as image and text. 
The performance of existing methods usually relies on labeled data for model training. However, cross-media data is very labor consuming to collect and label, so how to transfer valuable knowledge in \emph{\textbf{existing data}} to \emph{\textbf{new data}} is a key problem towards  application.
For achieving the goal, this paper proposes deep cross-media knowledge transfer (DCKT) approach, which transfers knowledge from a large-scale cross-media dataset to promote the model training on another small-scale cross-media dataset. The main contributions of DCKT are: 
(1) \emph{\textbf{Two-level transfer architecture}} is proposed to jointly minimize the media-level and correlation-level domain discrepancies, which allows two important and complementary aspects of knowledge to be transferred: intra-media semantic and inter-media correlation knowledge. It can enrich the training information and boost the retrieval accuracy.
(2) \emph{\textbf{Progressive transfer mechanism}} is proposed to iteratively select training samples with ascending transfer difficulties, via the metric of cross-media domain consistency with adaptive feedback. It can drive the transfer process to gradually reduce vast cross-media domain discrepancy, so as to enhance the robustness of model training.
For verifying the effectiveness of DCKT, we take the large-scale dataset XMediaNet as source domain, and 3 widely-used datasets as target domain for cross-media retrieval. Experimental results show that DCKT achieves promising improvement on retrieval accuracy.

\end{abstract}

%%%%%%%%% BODY TEXT
\section{Introduction}

	With the rapid development of computer and digital transition technology,  multimedia data such as image, text, video and audio can be found everywhere and exists as a whole to reshape our lives. 
	Human can naturally receive information from different sensory channels, such as vision and auditory. However, it has been indicated that the importances of sensory channels differ among people, resulting in different learning styles\cite{Visual}. For example, when students take in information with all the senses, such as seeing pictures and reading texts, they will have the highest efficiency of studying \cite{Visual}. If relevant multimedia data can be conveniently retrieved and provided, it will be very helpful to increase the efficiency of information acquisition for human.
	
\begin{figure}[t]
		\centering
		\begin{minipage}[c]{\linewidth}
			\centering
			\includegraphics[width=\textwidth]{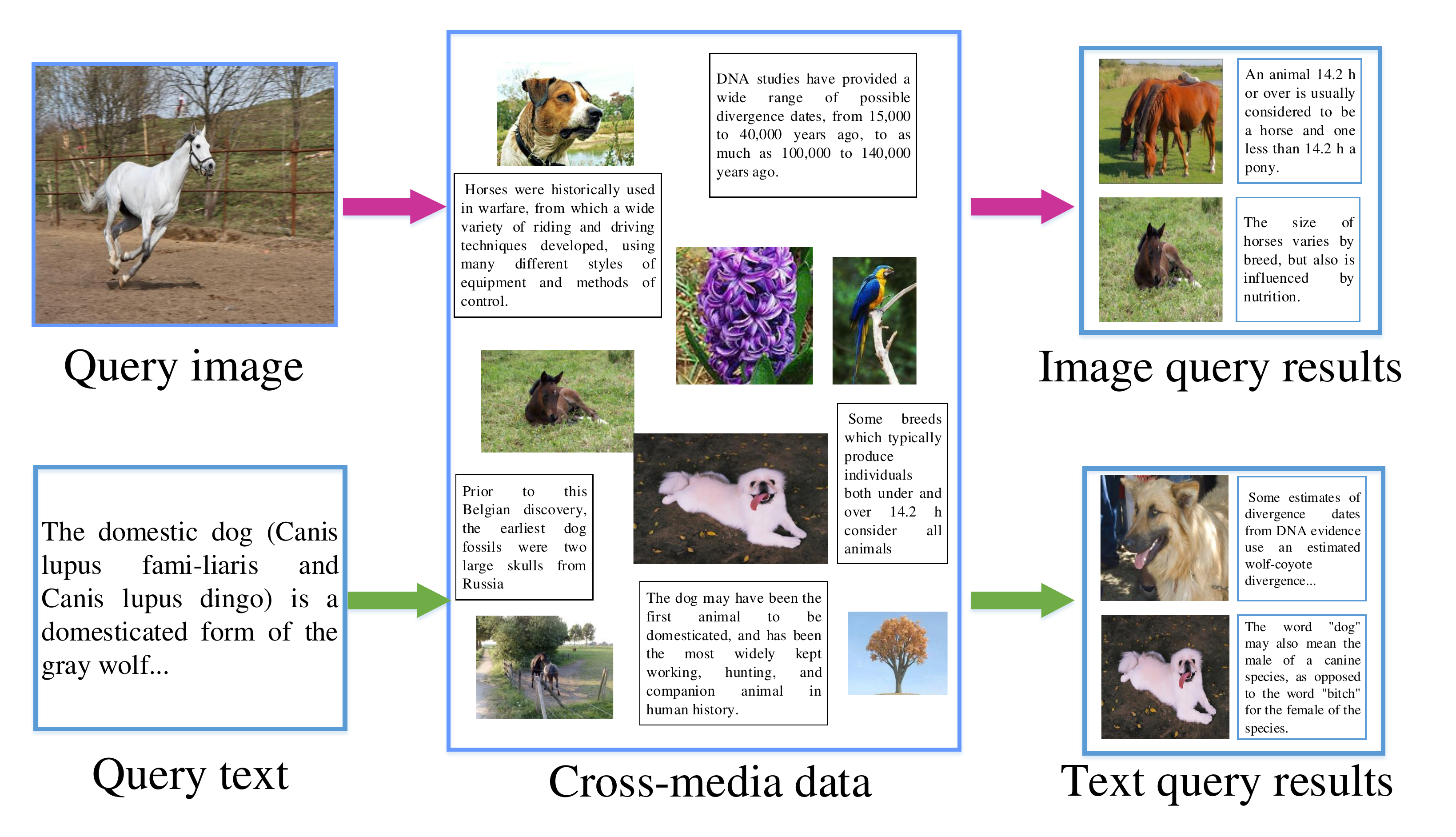}
		\end{minipage}%
				\setlength{\abovecaptionskip}{0.2cm}
		\caption{An example of cross-media retrieval.}\label{fig:cross-media}
	\end{figure}

	Cross-media retrieval \cite{DBLP:journals/corr/PengHZ17a} is such a kind of technique to flexibly provide data of different media types, with one query of any media type. Figure \ref{fig:cross-media} shows an example of cross-media retrieval, which includes two media types: image and text. As a highlighting research hotspot, cross-media retrieval has the advantage for realizing the coordination of different media types compared with traditional single-media retrieval. To perform cross-media retrieval, we have to deal with  ``heterogeneity gap". This means that different media types have inconsistent representation forms, so the similarity of them cannot be directly measured in their original feature spaces.
	Intuitively, the mainstream methods of cross-media retrieval are common representation learning, which aim to project data of different media types into an intermediate common space \cite {HotelingBiometrika36RelationBetweenTwoVariates, RasiwasiaMM10SemanticCCA, ZhaiTCSVT2014JRL, DBLP:conf/ijcai/PengHQ16}. Among them, deep neural network (DNN) based methods have currently become an active topic, which take DNN as basic model to perform common representation projection
	\cite{ngiam32011multimodal,srivastava2012learning,feng12014cross,DBLP:conf/ijcai/PengHQ16,DBLP:journals/corr/HuangPY17}.
	
	   Cross-media retrieval is still a challenging problem, and the performance of existing methods usually relies on labeled data for model training. However, insufficient training data is a common and severe challenge, especially for DNN-based methods. From the view of \emph{model training requirement}, because cross-media correlation is very complex and diverse, high-quality labeled data is crucial to provide cues for training ``good" DNN models. Insufficient data limits the training performance and easily leads to overfitting. From the view of \emph{human labor}, it is extremely labor-consuming to collect and label cross-media data. For example, if we want to collect data for ``water",  we need to see the images, read the texts, watch the videos, and even listen to the audio, and carefully judge whether the data is actually relevant to each other.
	   
	   In this situation, the idea of transfer learning \cite{DBLP:journals/tkde/PanY10, DBLP:conf/icml/LongC0J15, DBLP:conf/nips/LongZ0J16} becomes significant, which exploits general knowledge from source domain (usually a large-scale dataset) for relieving the problem of insufficient data. As known, cross-media data is quite labor consuming to collect and label, so existing labeled cross-media data is precious and valuable. It is a key problem towards application to distill knowledge from \emph{\textbf{existing data}} for boosting retrieval performance on \emph{\textbf{new data}}.
	   Nevertheless, existing transfer methods pay little attention to transfer between a large-scale cross-media dataset and a small-scale one. They also usually assume the domains share the same label space, which is often not satisfied due to the challenge of collecting cross-media data with the same semantic across domains.
	   So we consider the following problem: \emph{\textbf{How can we fully transfer knowledge from a large-scale cross-media dataset to promote the model training on another small-scale dataset, where they may have  different label spaces?}} For addressing this problem, this paper proposes deep cross-media knowledge transfer (DCKT) approach. 
		The main contributions of DCKT can be summarized as follows: 
		\begin{itemize}
		
		\item{\emph{\textbf{Two-level transfer architecture}} is proposed to jointly minimize the media-level and correlation-level domain discrepancies, which allows two important and complementary aspects of knowledge to be transferred: intra-media semantic and inter-media correlation knowledge. It can enrich the training information and boost the retrieval accuracy on target domain.}
		\item{ \emph{\textbf{Progressive transfer mechanism}} is proposed to iteratively select training samples with ascending transfer difficulties in target domain, via the metric of cross-media domain consistency with adaptive feedback. % It drives the cross-media transfer process with ascending difficulties,
		It can gradually reduce the vast cross-media domain discrepancy to enhance the robustness of model training.}
			
		\end{itemize}

		For performing knowledge transfer, a high-quality source domain is indispensable.
		In the experiment, we take a large-scale dataset XMediaNet as source domain, containing more than 100,000 labeled data with 200 distinct semantic categories. 
		For target domain, we adopt 3 widely-used datasets: Wikipedia, NUS-WIDE-10k and Pascal Sentences. Experimental results show that DCKT achieves promising improvement on cross-media retrieval accuracy.
		
		The following sections are organized as follows: Section \ref{sec:RelatedWork} gives a brief review of related work. Section \ref{sec:Architecture} presents the network architecture of DCKT, and Section \ref{sec:Opt} introduces the progressive transfer mechanism of DCKT. The experimental results and discussion are presented in Section \ref{sec:Experiment}, and finally Section \ref{sec:Conclusion} concludes this paper. 
		
\section{Related Work}
\label{sec:RelatedWork}
		\subsection{Cross-media retrieval}
		The current mainstream of cross-media retrieval is common representation learning, and the existing methods can be summarized as two main categories: shallow learning methods and DNN-based methods. Shallow learning methods usually take linear projections to convert cross-media data to common representation. A representative method is canonical correlation analysis (CCA) \cite{HotelingBiometrika36RelationBetweenTwoVariates}, which is a classical solution and extended by following works as \cite{RasiwasiaMM10SemanticCCA,DBLP:conf/iccv/RanjanRJ15}. Besides CCA, there are also many methods which incorporate various information to learn projection matrices as \cite{LiMM03CFA,ZhaiTCSVT2014JRL,DBLP:journals/tmm/KangXLXP15,DBLP:journals/tmm/HuaWLCH16}. Furthermore, link information can also be an important source of cross-media correlation, which has been used for clustering heterogeneous social media objects \cite{DBLP:conf/wsdm/QiAH12}.
		
		DNN-based cross-media retrieval methods are the currently active direction \cite{ngiam32011multimodal,kim2012learning,ICML2013DCCA,DBLP:conf/cvpr/YanM15,DBLP:conf/ijcai/PengHQ16,DBLP:journals/tcyb/WeiZLWLZY17}. Bimodal deep autoencoder \cite{ngiam32011multimodal} is a representative method, which is an extension of restricted Boltzmann machine (RBM). It can be seen as two autoencoders sharing the same code layer, where the common representation is obtained. Deep canonical correlation analysis (DCCA) \cite{ICML2013DCCA,DBLP:conf/cvpr/YanM15} is a non-linear extension of CCA, which can learn the complex non-linear transformations for two modalities. 
		Cross-media multiple deep networks (CMDN) \cite{DBLP:conf/ijcai/PengHQ16} jointly preserve the intra-media and inter-media information and then hierarchically combine them for improving the retrieval accuracy. 
		
		However, insufficient training data is a common and severe problem for existing methods. Inspired by the common use of large-scale single-media datasets like ImageNet \cite{ImageNet2012}, we intend to address this problem by exploiting a large-scale cross-media dataset \emph{XMediaNet} with general knowledge and transfer knowledge from it. This is useful towards real-world application where it is usually very hard to collect and label enough cross-media data.
								\begin{figure*}[t]
									\centering
									\begin{minipage}[c]{0.92\linewidth}
										\centering
										\includegraphics[width=0.98\textwidth]{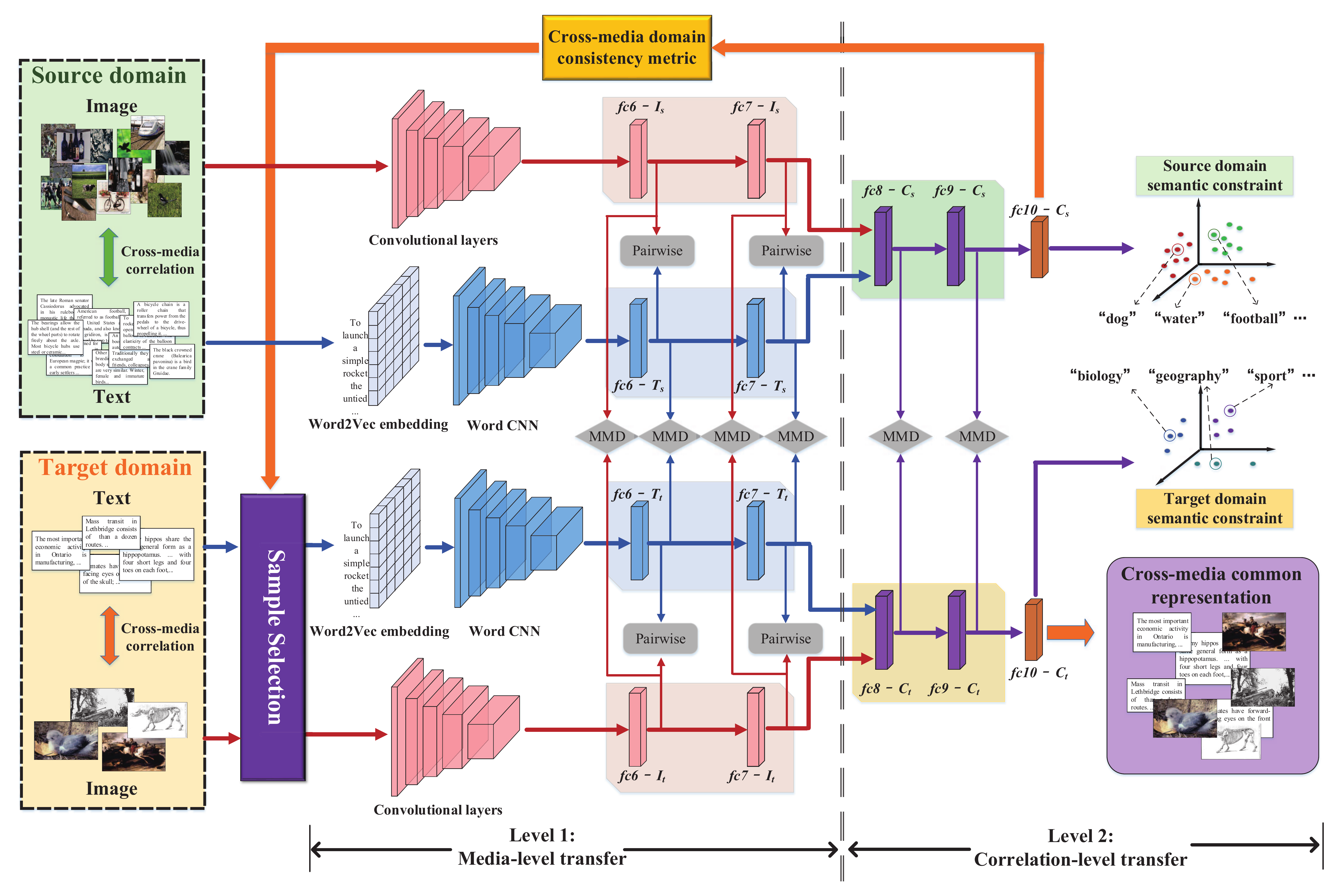}
									\end{minipage}%
									\setlength{\abovecaptionskip}{0.2cm}
									\caption{The overview of proposed deep cross-media knowledge transfer (DCKT) approach.}\label{fig:network}
								\end{figure*}	
		\subsection{Transfer Learning}
		It is natural that human can adapt the knowledge from already learned tasks to new tasks. Transfer learning \cite{DBLP:journals/tkde/PanY10} aims to simulate such mechanism, and relieve the problem of insufficient training data for a specific task.
		The focus of transfer learning is to reduce the domain discrepancy, which is widely used in DNN-based methods  \cite{DBLP:conf/nips/KrizhevskySH12,DBLP:conf/icml/LongC0J15, DBLP:conf/nips/LongZ0J16} for relieving the problem of insufficient training data, but mainly deals with single-media scenario. Besides, some works are proposed to perform transfer between different feature spaces  \cite{DBLP:conf/cvpr/TsaiYW16,DBLP:journals/tcyb/zhang2017semi} and multimedia domains \cite{DBLP:journals/tmm/YangZX15}. Transitive hashing network \cite{DBLP:journals/corr/CaoL016a} is proposed
		to learn from an auxiliary cross-media dataset to bridge two separate single-media datasets. Some works as \cite{DBLP:conf/www/QiAH11,DBLP:conf/mm/ShuQTW15} also propose to effectively transfer knowledge from text to image.
		Besides, Cross-media hybrid transfer network (CHTN) \cite{DBLP:journals/corr/HuangPY17} aims to transfer from a single-media source domain to cross-media target domain. 
		Different from the above works, this paper aims at transferring from a source domain with large-scale cross-media dataset to a target domain with small-scale cross-media dataset, where the label spaces are different. It is a challenging task because the intra-media semantic information, inter-media intrinsic  correlation and vast domain discrepancy should be jointly considered.

		\subsection{Curriculum Learning}
		The idea of progressive learning in this paper is inspired by curriculum learning (CL). The motivation of CL is simple: to first learn from easy samples, and gradually learn from harder samples \cite{DBLP:conf/icml/BengioLCW09}, which aims to reduce the negative effects brought by noisy data in early period of training. It can be also applied for deciding learning order of tasks \cite{DBLP:conf/cvpr/PentinaSL15}. Self-paced learning (SPL) is based on CL, which designs a weighted loss term on all samples in the learning objective \cite{DBLP:conf/nips/KumarPK10}, and can be regarded as CL's implementation as indicated in \cite{DBLP:journals/tip/GongTMLKY16}. 
		CL has been applied in many problems like image classification \cite{DBLP:journals/tip/GongTMLKY16} and object tracking \cite{DBLP:conf/cvpr/SupancicR13}.

		This paper adopts the idea of CL to assign samples with different transfer difficulties by metric of cross-media domain consistency. This is an iterative process with adaptive feedback, which gradually reduces the discrepancy between cross-media domains to enhance the robustness of model training, and improve retrieval accuracy on cross-media target domain.

		\section{Network Architecture of DCKT}
		\label{sec:Architecture}

		This section will introduce the network architecture of DCKT in Figure \ref{fig:network}. The training process of progressive transfer, including the domain consistency metric and sample selection, will be further introduced in Section \ref{sec:Opt}.
		
		This paper focuses on the scenarios where source and target domains both have two media types (i.e., image and text), but DCKT can be simply extended to more than two media types by adding pathways. 
		The end-to-end architecture of DCKT can be seen as two levels: media-level transfer and correlation-level transfer. 
		We denote the source domain as \emph{Src}$=\{(i_{s}^p,t_{s}^p),y^{p}_{s}\}_{p=1}^{P}$, where $(i_{s}^p, t_{s}^p)$ is the $p$-th image/text pair with label $y^{p}_{s}$. Similarly, the target domain includes training set \emph{$Tar_{tr}$}$=\{(i_{t}^q,t_{t}^q),y^{q}_{t}\}_{q=1}^{Q}$ and testing set \emph{$Tar_{te}$}$=\{(i_{t}^m,t_{t}^m)\}_{m=1}^{M}$. The aim of DCKT is to exploit both \emph{Src} and \emph{$Tar_{tr}$} to train the model for generating common representation of \emph{$Tar_{te}$}, which is $c_{t}(I)^m$ and  $c_{t}(T)^m$ for each image and text. After this, the cross-media retrieval can be performed by distance computing with common representation.

		\subsection{Level 1: Media-level Transfer}
		As the two domains both have two media types, the domain discrepancy can come from two aspects: (1) \emph{\textbf{Media-level discrepancy}}, which means the intra-media semantic information in two domains has discrepancy; (2) \emph{\textbf{Correlation-level discrepancy}}, which means the inter-media correlation information in two domains has discrepancy. Media-level transfer aims to address the media-level discrepancy by feature adaptation of the same media type between two domains. 
		
		For each domain, we have two pathways for image and text respectively, and the two domains have the same architecture.	For image pathway, we take widely-used VGG19 \cite{DBLP:journals/corr/SimonyanZ14a} as basic model. We keep all the layers of VGG19 except the last fully-connected layer, % which is corresponding to ImageNet \cite{ImageNet2012} of ImageNet large-scale visual recognition challenge (ILSVRC).
		and each input image is converted to 4,096-d representations via $fc6-I_s$/$fc7-I_s$ for source domain, and $fc6-I_t$/$fc7-I_t$ for target domain. For text pathway, we first embed each word into a vector via Word2Vec model \cite{DBLP:conf/nips/MikolovSCCD13}, and then generate the 300-d input feature vector of each text with Word CNN\cite{DBLP:conf/emnlp/Kim14}. Similar to image pathway, the input text feature will pass through two fully connected layers, namely $fc6-T_s$/$fc7-T_s$ and $fc6-T_t$/$fc7-T_t$.
				
		Between the two domains, we achieve media-level transfer by feature adaptation \cite{DBLP:conf/icml/LongC0J15} via minimizing the maximum mean discrepancy (MMD) \cite{gretton2012kernel} of the same media type. Taking image as an example, we use $I_s = \{i_s\}$ and $I_t =\{i_t\}$ to denote the distributions of images in \emph{$Src$} and \emph{$Tar_{tr}$}. 
		$\mu_k(a)$ denotes the mean embedding of $a$ in reproducing kernel Hibert space (RKHS) $\mathcal{H}_k$, and 
		$\mathbf{E}_{\textsc{x}\sim a}f(\textsc{x})=\left \langle f(\textsc{x}),\mu _k(a) \right \rangle_{\mathcal{H}_k}$ for $f\in \mathcal{H}_k$. 
		So the squared MMD $m_k^2(I_s,I_t)$ is denoted as follows:  		
		\begin{align}
		\label{eq:MMD}
			m_k^2(I_s,I_t)\overset{\Delta}{=}\left \| \mathbf{E}_{I_s}[\phi(i_{s},\theta_{I_s})]-\mathbf{E}_{I_t}[\phi(i_{t},\theta_{I_t})] \right \|^2_{\mathcal{H}_k}
		\end{align}
		where $\phi$ denotes a network layer's output, and $\theta_x$ denotes the network parameters for each pathway. For example, $\theta_{I_s}$ means parameters of \textbf{I}mage pathway in \textbf{s}ource domain, and $\theta_{I_t}$ means those of \textbf{I}mage pathway in \textbf{t}arget domain.
		%\begin{align}
		%	m_k^2(T_s,T_t)\overset{\Delta}{=}\left \| \mathbf{E}_{T_s}[\phi(t_{s},\theta_{Ts})]-\mathbf{E}_{T_s}[\phi(t_{t},\theta_{Tt})] \right \|^2_{\mathcal{H}_k}
		%\end{align}
		
		MMD is computed in a layer-wise style, which is between the corresponding layers of two domains, i.e., $fc6-I_s$/$fc6-I_t$, $fc7-I_s$/$fc7-I_t$ for image, and $fc6-T_s$/$fc6-T_t$, $fc7-T_s$/$fc7-T_t$ for text. By minimizing MMD, the media-level domain discrepancy can be reduced, which can align the single-media representation of two domains for knowledge transfer. The \emph{MMD loss} functions of image and text can be defined as: 
		\begin{align}
				Loss_{MMD_I} = \sum_{l=l_6}^{l_7}m_k^2(I_s,I_t) \\
				Loss_{MMD_T} = \sum_{l=l_6}^{l_7}m_k^2(T_s,T_t)
		\end{align}
		where $Loss_{MMD_I}$ and $Loss_{MMD_T}$ mean MMD loss functions for two media types.					
		
		Besides, in two domains, each pair of image and text as $(i_{s}^p,t_{s}^p)$ and $(i_{t}^q,t_{t}^q)$ exists together to represent closely relevant semantic, which is an important coexistence cue for cross-media retrieval. We preserve such pairwise constraint during transfer process via reducing the representation difference of each pair, which is a commonly-used criterion in cross-media retrieval \cite{feng12014cross,DBLP:journals/corr/HuangPY17}. Specifically, we use Euclidean distance as measurement, denoted as: 
		\begin{align}
			d^2(i^p_s,t^p_s)=\left \| \phi(i^p_s,\theta_{I_s})-\phi(t^p_s, \theta_{T_s}) \right \|^2 \\
			d^2(i^q_t,t^q_t)=\left \| \phi(i^q_t,\theta_{I_t})-\phi(t^q_t, \theta_{T_t}) \right \|^2
		\end{align} 
		Similar to what we have in Equation \ref{eq:MMD}, $\theta_x$ denotes the network parameters for each pathway. 
		Then we get the \emph{pairwise constraint loss} for two domains as:
		\begin{align}
			Loss_{Pair_s} = \sum_{l=l_6}^{l_7}\sum_{p=1}^{P}{d^2(i^p_s,t^p_s)} \\
			Loss_{Pair_t} = \sum_{l=l_6}^{l_7}\sum_{q=1}^{Q}{d^2(i^q_t,t^q_t)}
		\end{align}
		where $Loss_{Pair_s}$ and $Loss_{Pair_t}$ mean pairwise constraint loss for two domains, which are also computed in a layer-wise style between corresponding layers $fc6-I_s$/$fc6-T_s$, $fc7-I_s$/$fc7-T_s$ for source domain, and $fc6-I_t$/$fc6-T_t$, $fc7-I_t$/$fc7-T_t$ for target domain.
		By minimizing the MMD loss and pairwise constraint loss, we can transfer the intra-media semantic information from source domain to target domain, as well as avoiding damaging the data-coexistence relationship.

		\subsection{Level 2: Correlation-level Transfer}
				
		Cross-media domain discrepancy not only lies in the difference within each media type, but also in the correlation patterns for them to be correlated with each other.
		Correlation-level transfer aims to align the inter-media correlation of the two domains. For capturing the cross-media correlation in each domain, we adopt the strategy of shared layers to generate the common representation for different media types as \cite{DBLP:journals/corr/HuangPY17}. 		
		
		In the two domains, both image and text pathways will share two fully-connected layers. So the parameters of shared layers can fit the semantic learning of both two media types, which has the ability to capture inter-media correlation.
		%Now that the two shared fully-connected layers can capture the inter-media correlation information in two domains, 
		We add MMD loss function between the shared layers for correlation-level transfer. Similar to media-level transfer, we compute the \emph{MMD loss} function as follows:
		\begin{align}
			Loss_{MMD_C} = \sum_{l=l_8}^{l_9}m_k^2(C_s,C_t)
		\end{align}			
		where $l_{8/9}$ means the corresponding shared layers in two domains, i.e., $fc8-C_s$/$fc8-C_t$ and $fc9-C_s$/$fc9-C_t$ in Figure \ref{fig:network}, and $C_s$ and $C_t$ mean the output of shared layers of two domains.
		By minimizing $Loss_{MMD_C}$, the correlation-level domain discrepancy can be reduced, which aligns the inter-media correlation of two domains for knowledge transfer. 
		
		Besides, we should preserve the semantic information to maintain the semantically discriminative ability of common representation. This is intuitively achieved by semantic constraints with \emph{semantic loss} functions as follows:
		\begin{align}
				Loss_{Se_s} = \sum_{p=1}^{P} ( f_{sm}(i_s^p, y_s^p, \theta_{C_s}) + f_{sm}(t_s^p, y_s^p, \theta_{C_s}))\\
				Loss_{Se_t} = \sum_{q=1}^{Q} ( f_{sm}(i_t^q, y_t^q, \theta_{C_t}) + f_{sm}(t_t^q, y_t^q, \theta_{C_t}))
		\end{align}
		where $\theta_{C_s}$ and $\theta_{C_t}$ are the network parameters for pathways of \textbf{s}ource and \textbf{t}arget domains, and $f_{sm}$ is the softmax loss function. 		
		
		The architecture of DCKT is end-to-end, so the two levels of transfer can be jointly performed to mutually boost. It comprehensively allows the knowledge from cross-media source domain to be propagated to target domain. In this way, DCKT can enrich the training information with supplementary information of both intra-media semantic and inter-media correlation knowledge, thus promoting the model training performance and improve retrieval accuracy.
				
		\section{Progressive Transfer Mechanism}
		\label{sec:Opt}		
				
		All the introduced loss functions are able to be minimized by Stochastic Gradient Descent (SGD), so DCKT can be simply trained by simultaneously optimizing all of them with all data in $Src$ and $Tar_{tr}$ as input. However, because the discrepancy of two cross-media domains is usually quite vast with different label spaces, it may bring much noise and mislead the model training, especially for ``empty" models. So we propose a progressive transfer mechanism to gradually reduce the cross-media domain discrepancy.
		%However, cross-media domains can be viewed as a composition of information, which represent not only intra-media but also inter-media information, so the domain discrepancy is harder to directly reduce than single-media domains. Also, because the two domains have different label spaces, 
		%Also, because cross-media data is very hard to collect, it is common that we cannot collect large-scale data of the same label spaces for target domain, so that the source and target domains have different label spaces. It may bring much noise and confuse the model if we train them from ``empty" model.
		
\begin{figure}[t]
					\centering
					\begin{minipage}[c]{\linewidth}
						\centering
						\includegraphics[width=0.85\textwidth]{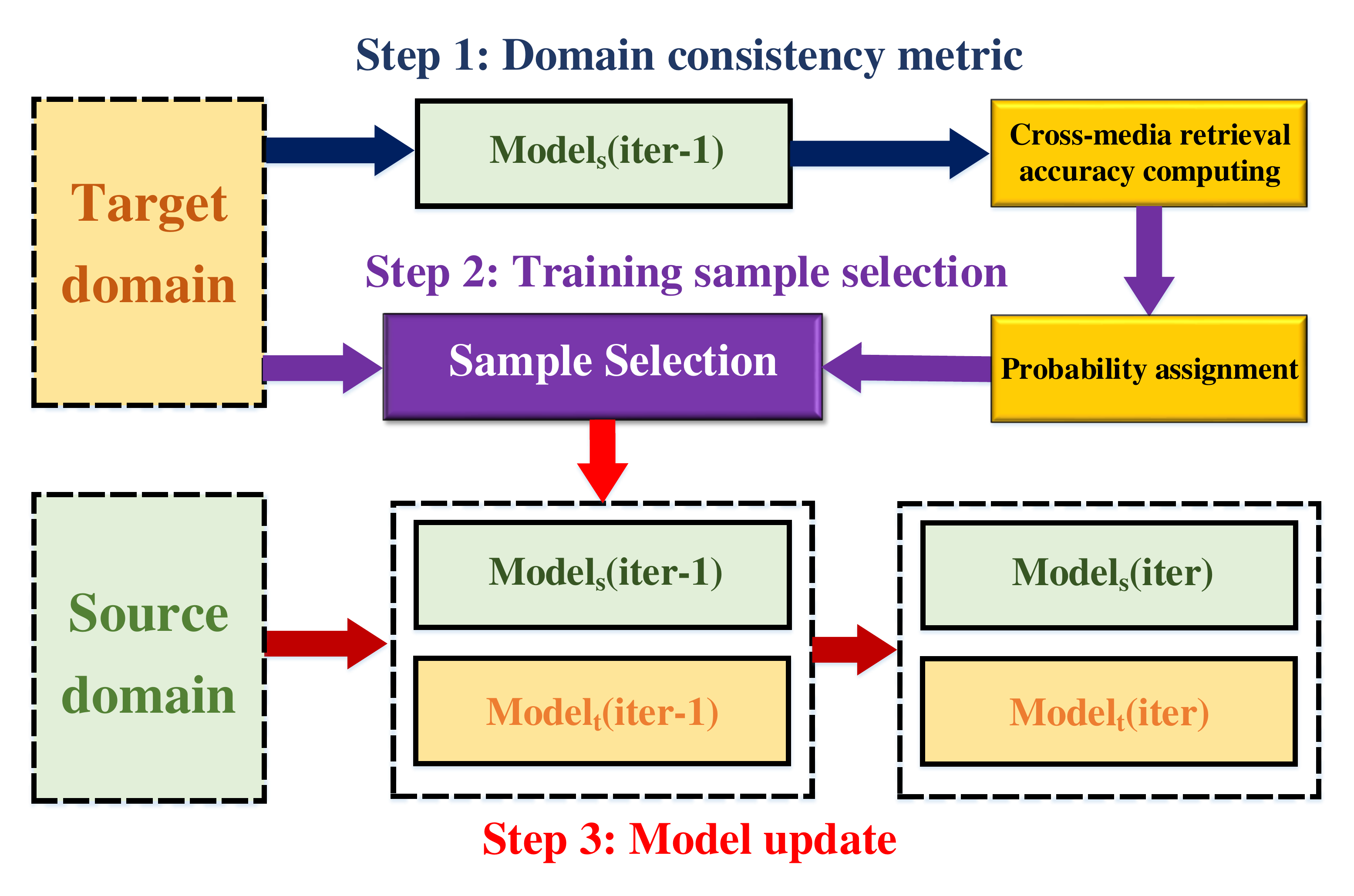}
					\end{minipage}%
					\setlength{\abovecaptionskip}{0.2cm}
					\caption{Process in each iteration of progressive transfer.}\label{fig:iter}
					\vspace{-2.8mm}
				\end{figure}

		To start from a ``safe" point, we first pre-train the model for each domain separately, removing all the MMD loss linking the two domains. For convenience, we denote the networks for two domains as $Model_s$ and $Model_t$.
		Then we progressively transfer the knowledge from source domain to target domain, which is an iterative process shown as Figure \ref{fig:iter}. Because source domain is relatively large-scale and reliable, we take $Model_s$ as reference model to perform sample selection in target domain. The motivation is intuitive: 
		In early period of training, we choose ``easy" samples in $Tar_{tr}$ whose cross-media correlation can be successful molded by $Model_s$, which are of high consistency with source domain. For example, although the label spaces are different, some categories such as ``sport" and ``football" have strong consistency. In late period of training when the model is stable, we can incorporate ``harder" samples with low domain consistency to further adapt to target domain.
		
		In each iteration $iter$, we generate common representation (class probability vector) for $Tar_{tr}$ as $C_s$ by $Model_s(iter)$, including $C_s(I)$ and $C_s(T)$. Next, we perform bi-directional cross-media retrieval and evaluate domain consistency according to the accuracy, which is Image$\rightarrow$Text and Text$\rightarrow$Image. Taking Image$\rightarrow$Text as an example,
		we compute the cosine distance between each image $c_s(I)^q$ and every text in $C_s(T)$, and then rank them to get  the AP score of $c_s(I)^q$ as:
				\begin{align}
				\label{eq:Ap}
						AP(I)^q = \frac 1 R \sum_{k=1}^Q \frac {R_k} k \times rel_k
				\end{align}
		where $R$ is the number of text with the same label of $c_s(I)^q$, $R_k$ is the number of relevant text in top-$k$ results. $rel_k$ indicates whether $c_s(I)^q$ and $k$-th result have the same label.
		
		A high $AP(I)^q$ means $Model_s(iter)$ successfully captures the cross-media correlation of $i^q_t$, i.e., the source domain contains closely relevant knowledge of $i^q_t$, so it can be regarded as an ``easy" transfer sample. Similarly we have $AP(T)^q$ and obtain:
		\begin{align}
		\label{eq:apq}
			AP^q = AP(I)^q + AP(T)^q
		\end{align}		
		where $AP^q$ can be used to estimate domain consistency of a pair $(i^q_t, t^q_t)$. During the training process, $Model_s$ is also iteratively updated, so $AP^q$ should be computed in each iteration. A high $AP^q$ means $q$-th pair is proper to be a bridge of the two domains. We assign the probability to be selected for each pair as:
		\begin{align}
		\label{eq:Prob}		
		 Prob(q) = \alpha[1- \log_2(\frac {max(AP)- AP^q} {max(AP)\times iter} +1)]
		\end{align}
		where $max(AP)$ is the maximal value of $AP^q$, and $\alpha \in (0,1]$ is the upper bound of $Prob(q)$. $\alpha$ prevents the ``easiest" samples from always being selected, which leads to the risk of overfitting. When $iter$ increases, the value of item $({max(AP)- AP^q}) / ({max(AP)\times iter)}$ will turn small, which means the selection will gradually become random sampling. The above process can be summarized as Algorithm \ref{alg:Ptransfer}.

\begin{algorithm}[t]
\small
\caption{: Progressive Transfer}
 \label{alg:Ptransfer}
  \begin{algorithmic}[1]
  \REQUIRE
  Training data $Src$ and $Tar_{tr}$, maximal iteration number $MI$, and epoch number in each iteration $Ep$.
  \STATE Pre-train $Model_s$ and $Model_t$ separately, denoted as $Model_s(0)$ and $Model_t(0)$. Set $iter=1$.
  \REPEAT
  \STATE Generate the common representation for $Tar_{tr}$ with $Model_s(iter-1)$ as $C_s(I)$ and $C_s(T)$.
  \STATE Compute $AP(I)^q$ and $AP(T)^q$ for all $(i^q_t, t^q_t) \in Tar_{tr}$.
  \STATE Compute $AP^q$ via Equation \ref{eq:apq}.
  \STATE Estimate $Prob(q)$ via Equation \ref{eq:Prob}, and select training sample set $Tar_{tr}(iter)$.
  \STATE Train model for $Ep$ epochs with $Src$ and $Tar_{tr}(iter)$, to get $Model_s(iter)$ and $Model_t(iter)$.
    \STATE $iter=iter+1$.
  \UNTIL $iter=MI$.
  \RETURN $Model_s(MI)$ and $Model_t(MI)$.
  \end{algorithmic}
\end{algorithm}

	After training, each testing data can be converted as common representation (actually class probability vector), and then the cross-media retrieval can be performed by distance metric. Note that in testing stage, the image and text data can be input separately, whose labels and pairwise correlation are not used at all. This setting is widely adopted in cross-media retrieval as \cite{DBLP:journals/tcyb/WeiZLWLZY17,DBLP:journals/corr/HuangPY17}.

		\section{Experiments}
		\label{sec:Experiment}
						
		\subsection{Details of Implementation}
		
		The architecture of DCKT is easy to implement, and the parts of two domains share the same architecture. For image we use VGG19 \cite{DBLP:journals/corr/SimonyanZ14a} as basic model to generate convolutional feature maps of pool5, which is pre-trained by ImageNet \cite{ImageNet2012} of ImageNet large-scale visual recognition challenge (ILSVRC) 2012. For text we first embed each word into a vector via Word2Vec model \cite{DBLP:conf/nips/MikolovSCCD13}, and then generate $300$-d text feature following \cite{DBLP:conf/emnlp/Kim14}. 
		The classification layers $fc10-C_s$ and $fc10-C_t$ are fully-connected layers of the same unit number with the semantic categories in each domain. All the other layers are fully-connected layers of $4,096$ units, including $fc6-I_{s/t}$, $fc7-I_{s/t}$, $fc6-T_{s/t}$, $fc7-T_{s/t}$, $fc8-C_{s/t}$, and $fc9-C_{s/t}$.
		The \emph{pairwise constraint loss} functions are implemented by contrastive loss layers from Caffe\footnote{http://caffe.berkeleyvision.org}. The \emph{MMD loss} functions are implemented following \cite{DBLP:conf/icml/LongC0J15}, by which the knowledge transfer of the two domains is actually performed. As for network parameters, we set the initial learning rates as 0.01, and the weight decay 0.0005. In the mechanism of progressive training in Algorithm \ref{alg:Ptransfer}, we set $\alpha$ as 0.2, $Ep$ as $1$, and $MI$ as 10. These parameters will be further analyzed in Section \ref{sec:Para}.

		\subsection{Datasets}
		\subsubsection{Source Domain}			
		To serve as the source domain, the dataset should be large-scale, high-quality, and of general knowledge like ImageNet \cite{ImageNet2012} and Google News corpus\cite{DBLP:journals/corr/abs-1301-3781}, so that the knowledge is proper to be adapted to other domains.	
		
		{\textbf {XMediaNet\cite{DBLP:journals/corr/PengHZ17a}} dataset is adopted to serve as the source domain. It is a large-scale dataset with 5 media types, which has more than 100,000 media instances of text, image, audio, video and 3D model. All the instances are manually collected and labeled from famous websites such as Wikipedia, Flickr, Youtube, Findsounds, Freesound, and Yobi3D. It includes 200 distinct semantic categories based on wordNet hierarchy to avoid semantic confusion, including 47 animal species like ``dog" and 153 artifact species like ``airplane". In this paper, we focus on the scenario of image and text, so we choose the training set of image and text data from XMediaNet with 32,000 pairs. %Some examples in XMediaNet are shown as Figure \ref{fig:xmn}.
	%	
	%	\begin{figure}[t]
	%		\centering
	%		\begin{minipage}[c]{\linewidth}
	%			\centering
	%			\includegraphics[width=0.95\textwidth]{xmn.pdf}
	%		\end{minipage}%
	%		\setlength{\abovecaptionskip}{0.2cm}
	%		\caption{Some examples of image and text in XMediaNet.}\label{fig:xmn}
	%		\vspace{-2mm}
	%	\end{figure}
								
		\subsubsection{Target Domain}		
		For target domain, we adopt 3 widely-used datasets to conduct cross-media retrieval, namely Wikipedia, NUS-WIDE-10k and Pascal Sentences. They all have two media types image and text. The dataset split is strictly according to  \cite{feng12014cross,DBLP:conf/ijcai/PengHQ16, DBLP:journals/corr/HuangPY17}, shown as Table \ref{table:dataset}.
		
				\begin{table}[htb]
				 \begin{center}
				 \resizebox{0.8\linewidth}{!}{
				 \begin{tabular}{|c|c|c|c|c|} 
				 \hline
				 \multirow{2}{*}{Dataset} & \multicolumn{3}{c|}{Split}\\
				 \cline{2-4}
				  & Training & Testing & Validation \\   \hline
				  \multirow{1}{*} 				  
				  Wikipedia \cite{RasiwasiaMM10SemanticCCA} & 2,173 & 462 & 231 \\
				   NUS-WIDE-10k \cite{NUSWIDE,feng12014cross} & 8,000 & 1,000 & 1,000 \\
				   Pascal Sentences \cite{PascalSentence} & 800 & 100 & 100 \\
				  \hline						
				 \end{tabular} 
				 }
				 \end{center}
				 \vspace{-2mm}
				 \caption{The size and split of each dataset as target domain.}
				 \label{table:dataset}
				 \end{table}							
						
		%{\textbf {Wikipedia dataset}} \cite{RasiwasiaMM10SemanticCCA} is the most popular dataset for evaluation of cross-media retrieval, which has been adopted by many previous works as  \cite{ZhaiTCSVT2014JRL,feng12014cross,DBLP:conf/ijcai/PengHQ16}. Wikipedia dataset contains 2,866 image/text pairs of 10 high-level semantic categories such as ``art" and ``history". The dataset is split as a training set of 2,173 pairs, a testing set of 462 pairs, and a validation set of 231 pairs \cite{feng12014cross,DBLP:conf/ijcai/PengHQ16}. 
						
		%{\textbf {NUS-WIDE-10k dataset}} \cite{feng12014cross} is a subset of NUS-WIDE dataset\cite{NUSWIDE}, which includes 10,000 images with corresponding text tags of 10 semantic categories. It is also randomly split into a training set of 8,000 pairs, a testing set of 1,000 pairs, and a validation set of 1,000 pairs. In each set, the data numbers of every categories are the same.

		%{\textbf {Pascal Sentences dataset}} \cite{PascalSentence} is constructed with 2008 PASCAL development kit. It has 1,000 images of 20 semantic categories, and each image is with 5 sentences. Pascal Sentence dataset is split into a training set of 800 pairs, a testing set of 100 pairs, and a validation set of 100 pairs. Similar to NUS-WIDE-10k dataset, the pars in each set are evenly from the 20 categories.
						
		\subsection{Compared Methods}
		We compare our proposed DCKT approach with totally 12 state-of-the-art methods with source codes from the authors of original papers, namely 
		CCA \cite{HotelingBiometrika36RelationBetweenTwoVariates}, CFA \cite{LiMM03CFA}, KCCA (with Gaussian kernel) \cite{DBLP:journals/neco/HardoonSS04}, Corr-AE \cite{feng12014cross}, JRL \cite{ZhaiTCSVT2014JRL}, LGCFL \cite{DBLP:journals/tmm/KangXLXP15}, DCCA \cite{DBLP:conf/cvpr/YanM15}, CMDN \cite{DBLP:conf/ijcai/PengHQ16} Deep-SM \cite{DBLP:journals/tcyb/WeiZLWLZY17}, CHTN \cite{DBLP:journals/corr/HuangPY17}, ACMR \cite{DBLP:conf/mm/WangYXHS17}, and CCL \cite{DBLP:journals/corr/PengQHY17}.
		
		Due to the wide range of comparison methods, their original papers adopt different input settings. For example, CHTN and Deep-SM are based on AlexNet and take original image pixels as input, while others like CCL take feature vectors as input. For fair comparison,  we replace the AlexNet of CHTN and Deep-SM with VGG19, and use the 4,096-d VGG19 image feature for methods which need feature vector as input. As for text, we use the same 300-d Word CNN text features for all the methods, which is the same with our DCKT.

		\subsection{Evaluation Metrics}
						
		We conduct cross-media retrieval task with two directions: Image$\rightarrow$Text and Text$\rightarrow$Image. Taking Image$\rightarrow$Text as an example, the retrieval process is conducted as follows: (1) Get the common representation for all images and texts in testing set. (2) Take one image as query, and compute the cosine distance between the common representation of query image and all texts. (3) Rank all the texts in testing set with similarities according to the distances.	
						
		The metric adopted for evaluating the retrieval results is mean average precision (MAP) score, which is the mean value of average precision (AP) scores of all queries. AP is computed as Equation \ref{eq:Ap}. \emph{All retrieval results} will be considered for the computation of MAP score following \cite{DBLP:journals/corr/PengQHY17,DBLP:journals/tcyb/WeiZLWLZY17,DBLP:journals/corr/HuangPY17}, instead of \emph{top-50 results} as \cite{feng12014cross,DBLP:conf/mm/WangYXHS17}.

 \begin{table}[htb]
 \begin{center}
 \scriptsize 
 \resizebox{\linewidth}{!}{
 \begin{tabular}{|c|c|c|c|c|} 
 \hline
 \multirow{2}{*}{Dataset}&
 \multirow{2}{*}{Method} & \multicolumn{3}{c|}{Task}\\
 \cline{3-5}
  & & Image$\rightarrow$Text & Text$\rightarrow$Image & Average \\
  \hline
  
  \multirow{13}{*}{\begin{tabular}{c} Wikipedia \\ dataset \end{tabular}} 
   &  \textbf{our DCKT} & \textbf{0.537} & \textbf{0.485} & \textbf{0.511} \\
   & CCL\cite{DBLP:journals/corr/PengQHY17}& 0.505 & 0.457 & 0.481 \\
   & ACMR\cite{DBLP:conf/mm/WangYXHS17}& 0.468 & 0.412 & 0.440 \\
   & CHTN\cite{DBLP:journals/corr/HuangPY17} & 0.523 & 0.460 & 0.492 \\
   & Deep-SM\cite{DBLP:journals/tcyb/WeiZLWLZY17}& 0.478 & 0.422 & 0.450 \\
   & CMDN\cite{DBLP:conf/ijcai/PengHQ16}& 0.487 & 0.427 & 0.457 \\
   & DCCA\cite{DBLP:conf/cvpr/YanM15}& 0.445 & 0.399 & 0.422 \\
   & LGCFL\cite{DBLP:journals/tmm/KangXLXP15}& 0.466 & 0.431 & 0.449 \\
   & JRL\cite{ZhaiTCSVT2014JRL}& 0.479 & 0.428 & 0.454 \\
   & Corr-AE\cite{feng12014cross}& 0.442 & 0.429 & 0.436 \\
   & KCCA\cite{DBLP:journals/neco/HardoonSS04}& 0.438 & 0.389 & 0.414 \\
   & CFA\cite{LiMM03CFA}& 0.319 & 0.316 & 0.318 \\
   & CCA\cite{HotelingBiometrika36RelationBetweenTwoVariates} & 0.298 & 0.273 & 0.286 \\
  \hline
 
  \multirow{13}{*}{\begin{tabular}{c} NUS-WIDE\\ -10k \\dataset \end{tabular}} 
  &  \textbf{our DCKT} & \textbf{0.556} & \textbf{0.584} & \textbf{0.570} \\
  & CCL\cite{DBLP:journals/corr/PengQHY17}& 0.481  & 0.520  & 0.501 \\
  & ACMR\cite{DBLP:conf/mm/WangYXHS17}& 0.519 & 0.542 & 0.531 \\
  & CHTN\cite{DBLP:journals/corr/HuangPY17}& 0.537  & 0.562  & 0.550 \\
  & Deep-SM\cite{DBLP:journals/tcyb/WeiZLWLZY17}& 0.497  & 0.478  & 0.488 \\
  & CMDN\cite{DBLP:conf/ijcai/PengHQ16}& 0.492  & 0.542  & 0.517 \\
  & DCCA\cite{DBLP:conf/cvpr/YanM15}& 0.452  & 0.465  & 0.459 \\
  & LGCFL\cite{DBLP:journals/tmm/KangXLXP15}& 0.453  & 0.485  & 0.469 \\
  & JRL\cite{ZhaiTCSVT2014JRL}& 0.466  & 0.499  & 0.483  \\
  & Corr-AE\cite{feng12014cross}& 0.441  & 0.494  & 0.468  \\
  & KCCA\cite{DBLP:journals/neco/HardoonSS04}& 0.351 & 0.356 & 0.354 \\
  & CFA\cite{LiMM03CFA}& 0.406 & 0.435 & 0.421 \\
  & CCA\cite{HotelingBiometrika36RelationBetweenTwoVariates}& 0.167 & 0.181 & 0.174 \\
  
  \hline
  
   \multirow{13}{*}{\begin{tabular}{c} Pascal \\ Sentences \\ dataset \end{tabular}}
  &  \textbf{our DCKT} & \textbf{0.582} & \textbf{0.587} & \textbf{0.585} \\
 & CCL\cite{DBLP:journals/corr/PengQHY17}& 0.576  & 0.561  & 0.569 \\ 
 & ACMR\cite{DBLP:conf/mm/WangYXHS17}& 0.538 & 0.544 & 0.541 \\ 
 & CHTN\cite{DBLP:journals/corr/HuangPY17}& 0.556  & 0.534  & 0.545 \\ 
 & Deep-SM\cite{DBLP:journals/tcyb/WeiZLWLZY17}& 0.560  & 0.539  & 0.550 \\ 
 & CMDN\cite{DBLP:conf/ijcai/PengHQ16}& 0.544  & 0.526  & 0.535 \\ 
 & DCCA\cite{DBLP:conf/cvpr/YanM15}& 0.568  & 0.509  & 0.539 \\ 
 & LGCFL\cite{DBLP:journals/tmm/KangXLXP15}& 0.539  & 0.503  & 0.521 \\ 
 & JRL\cite{ZhaiTCSVT2014JRL}& 0.563  & 0.505  & 0.534 \\ 
 & Corr-AE\cite{feng12014cross}& 0.532  & 0.521  & 0.527 \\ 
 & KCCA\cite{DBLP:journals/neco/HardoonSS04}& 0.488  & 0.446  & 0.467 \\ 
 & CFA\cite{LiMM03CFA}& 0.476  & 0.470  & 0.473 \\ 
 & CCA\cite{HotelingBiometrika36RelationBetweenTwoVariates}& 0.203  & 0.208  & 0.206 \\ 
   \hline
 \end{tabular} 
 }
 \end{center}
 \caption{MAP scores of our DCKT and compared methods. \emph{All retrieval results} are evaluated for comprehensive comparison, instead of \emph{top-50 results} as \cite{feng12014cross,DBLP:conf/mm/WangYXHS17}. }
 \label{table:Results}
 \vspace{-2mm}
 \end{table}

				\subsection{Experimental Results}
				\subsubsection{Comparison with State-of-the-art methods}
				Table \ref{table:Results} shows the retrieval accuracy of DCKT and compared methods. On Wikipedia dataset, DCKT gains the improvement from 0.492 to 0.511, compared with the method with highest MAP score CHTN. Among the compared methods, we can see that the shallow learning method JRL achieves comparable accuracy with DNN-based methods, and even outperforms Corr-AE, Deep-SM, and DCCA. This is probably because that the small scale of Wikipedia dataset is insufficient for deep network to get ideal training performance. On NUS-WIDE-10k and Pascal Sentences datasets, our DCKT achieves the best MAP scores, too. The above results show the stable advantage of DCKT compared with existing methods. This is because the two-level transfer network architecture and progressive transfer mechanism allow the intra-media semantic and inter-media correlation knowledge to be propagated to the target domain, improving training effectiveness on cross-media target domain.
				
				It should be noted that CHTN is also a transfer learning based method, which transfers knowledge from single-media source domain (ImageNet) to cross-media target domain. By comparing the MAP scores of DCKT and CHTN, it can be seen that it is helpful to transfer from a cross-media source domain, because the cross-media source domain has not only media-level knowledge, but also rich correlation-level knowledge. %Our DCKT keeps the best MAP scores on all 3 dataset, which shows that the two-level transfer network allows the semantic and correlation knowledge to be propagated to the target domain, which can jointly align the modality-specific representation and high-level semantic to minimize the domain discrepancy, and the  progressive transfer mechanism can drive the transfer processing with ascending difficulties, which gradually reduce the heterogeneity gap and domain discrepancy, improving training effectiveness on cross-media target domain.

				\subsubsection{Baseline Experiment}
				To further analyze the performance of DCKT, we conduct baseline experiments on 3 dataset. The results are shown in Table \ref{table:Baseline}. Due to the page limitation, we show the \textbf{average MAP scores} of retrieval in 2 directions.
				The basic idea of this paper is knowledge transfer, so the first question is: Is the knowledge transfer process actually helpful? To verify this, we perform retrieval with the separately pre-trained model of Wikipedia dataset, i.e., $Model_t(0)$. We denote the complete DCKT model as $DCKT_{Full}$. By comparing $Model_t(0)$ with $DCKT_{Full}$ in Table \ref{table:Baseline}, we can see that the transfer process achieves inspiring improvement.
				
				Then we verify the effectiveness of two key strategies of DCKT: \emph{Two-level transfer} and \emph{progressive transfer}. For \emph{two-level transfer}, we design 2 baselines: only with media-level transfer (Transfer\_1 in Table \ref{table:Baseline}) or correlation-level transfer (Transfer\_2 in Table \ref{table:Baseline}), and keep other parts unchanged. From Table \ref{table:Baseline} we can see that the combination of the two levels can achieve more improvement than either of them, which shows that the two levels of knowledge are complementary for cross-media retrieval.
				
				For \emph{progressive transfer}, we design 2 baselines: $DCKT_{All}$ means that in each iteration, we use all data in $Tar_{tr}$. $DCKT_{Random}$ means that we select samples randomly. It can be seen that although knowledge transfer is helpful, the domain discrepancy is vast in cross-media scenario, so $DCKT_{All}$ and $DCKT_{Random}$ both achieve lower MAP scores than $DCKT_{Full}$. We also observe that $DCKT_{Random}$ is slightly lower than ${DCKT_{All}}$, which is because that by arbitrary sampling, the model cannot have the whole view in each iteration, which brings negative effects than $DCKT_{All}$.
				
				Besides, there may exist category overlaps between the source and target domains. Wikipedia has no category overlap with XMediaNet dataset (0 of totally 10), while NUS-WIDE-10k has minor overlap (3 of 10), and Pascal Sentences has large overlap (12 of 20). $DCKT_{No\ overlap}$ means that we remove the overlap categories in XMediaNet dataset with NUS-WIDE-10k and Pascal Sentences datasets, respectively. The results are not sensitive to overlap, which shows our DCKT is robust for different label spaces.
							
				\begin{table}[htb]
				 \begin{center}
 				\resizebox{\linewidth}{!}{
				 \begin{tabular}{|c|c|c|c|c|} 
				 \hline
				 \multirow{2}{*}{Method} & \multicolumn{3}{c|}{Dataset}\\
				 \cline{2-4}
				  & Wikipedia & NUS-WIDE-10k & Pascal Sentences\\
				  \hline
				  
				  \multirow{7}{*}
				   \textbf{$DCKT_{Full}$} & \textbf{0.511} & \textbf{0.570} & \textbf{0.585} \\ 
				    $Model_t(0)$ & 0.459 & 0.527 & 0.529 \\ \hline \hline
				   Transfer\_1 & 0.491 & 0.555 & 0.565 \\
				   Transfer\_2 &  0.487 & 0.553 & 0.569 \\ \hline \hline
				   $DCKT_{All}$ & 0.498 & 0.560 & 0.574 \\
				   $DCKT_{Random}$ & 0.494 & 0.553 & 0.573 \\ \hline \hline
				   $DCKT_{No\ overlap}$ & -- & 0.566 & 0.579 \\
				  \hline			 
				 \end{tabular} 
				 }
				 \end{center}
				 \caption{Average MAP scores of baseline experiments.}
				 \label{table:Baseline}
				 \vspace{-2mm}
				 \end{table}		
				 
				\subsubsection{Parameter Analysis}
				\label{sec:Para}
				In this section we analyze the settings of parameters $MI$, $Ep$, and $\alpha$ in Algorithm \ref{alg:Ptransfer}. In our experiment, because the sizes of $Src$ and $Tar_{tr}$  are different, for ensuring in each iteration $Src$ can be processed throughout, we set $Ep = 1$ for it. Correspondingly, for $Tar_{tr}$ the epoch number is $P/Q$. As for $MI$, we set it as 10 in our experiment, and the performance will tend to be stable. They can also be intuitively adjusted according to validation set. 
				
				Next, $\alpha$ determines how many samples we can select in an iteration of progressive transfer. For investigating the impact of $\alpha$, we conduct DCKT with different $\alpha$ values. The impact is shown as Figure \ref{fig:alpha}. We can see that although we perform transfer based on pre-trained model $Model_t(0)$, the performance is seriously damaged with very small $\alpha$. When $\alpha$ increases, the MAP score will increase apparently until 0.2. Then the MAP scores are generally stable but tend to be lower. This shows that a large $\alpha$ means the ``easy" samples are always selected, which can lead to the risk of overfitting.
				
				\begin{figure}[t]
					\centering
						\begin{minipage}[c]{\linewidth}
							\centering
							\includegraphics[width=0.98\textwidth]{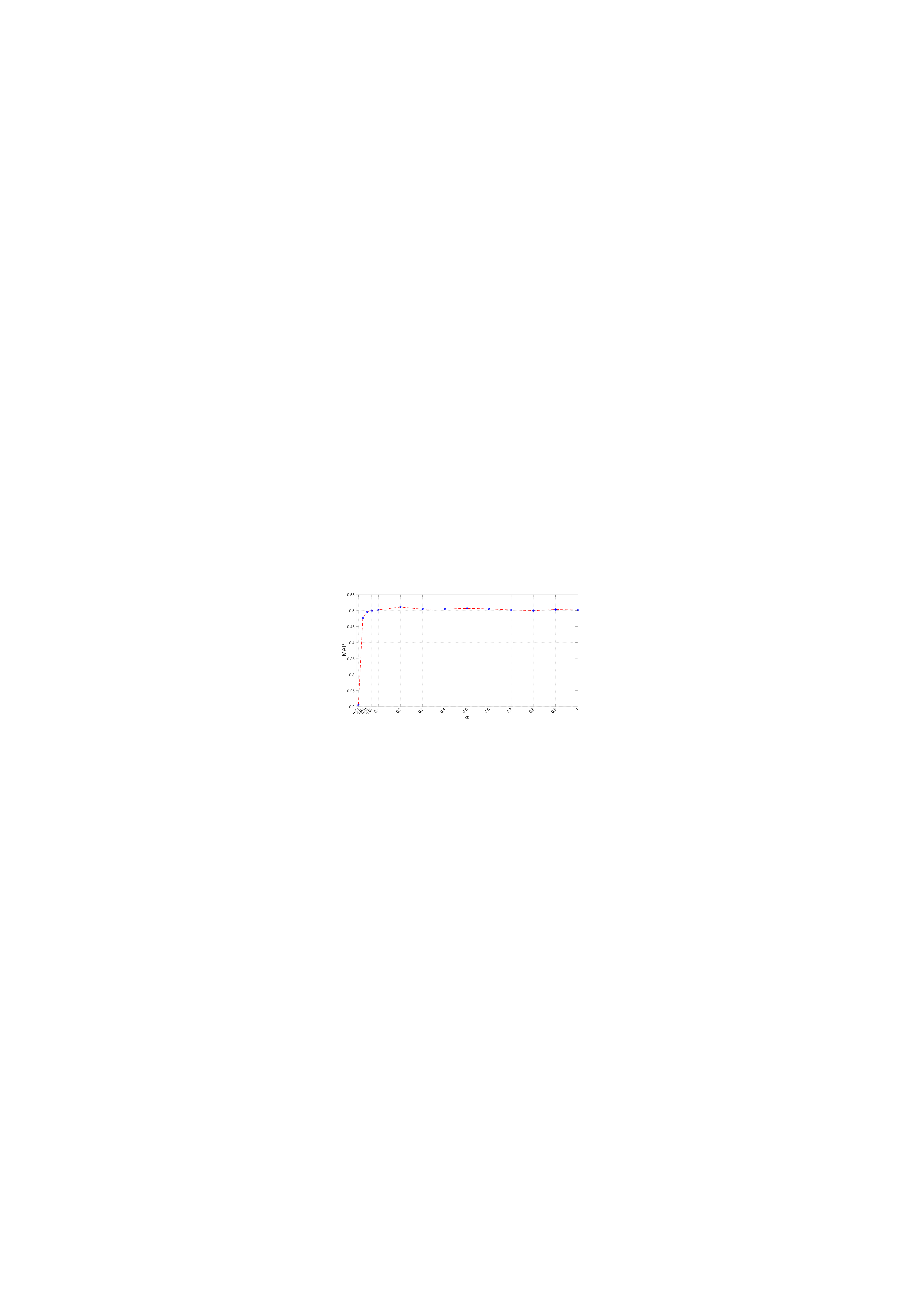}
						\end{minipage}%
						\setlength{\abovecaptionskip}{0.2cm}
						\caption{Impact of $\alpha$ on MAP score of Wikipedia dataset.}\label{fig:alpha}
						\vspace{-2mm}
				\end{figure}
				
			\section{Conclusion}
			\label{sec:Conclusion}
			This paper has proposed deep cross-media knowledge transfer (DCKT) approach, which transfers knowledge from a large labeled cross-media dataset as source domain to promote the performance of model training on target domain. 
			DCKT is a two-level transfer network to allow the intra-media and inter-media knowledge to be propagated to the target domain, which can enrich the training information and boost the retrieval accuracy on target domain. For addressing the vast domain gap, we propose progressive transfer mechanism to iteratively select training samples with ascending transfer difficulties in target domain, which can drive the cross-media transfer process to gradually reduce the vast cross-media domain discrepancy, and enhance the robustness.
			In the experiments, we take the large-scale dataset XMediaNet as source domain, and 3 widely-used datasets as target domain for cross-media retrieval. Experimental results show that DCKT achieves promising improvement on retrieval accuracy. For the future work, we intend to propose more effective strategy for sample selection, and extend DCKT for unsupervised transfer scenario, i.e, the semantic labels of target domain are unknown, which will further save the human labor of labeling data.

\section{Acknowledgments}
This work was supported by National Natural Science Foundation of China under Grants 61771025 and 61532005.

{\small
\bibliographystyle{ieee}
\bibliography{egbib}
}

\end{document}